\documentclass[aps,twocolumn,reprint]{revtex4-1}
\usepackage{amsmath,amssymb,amsthm,bm,braket,ascmac,bbm}

\usepackage{graphicx,xcolor} %chinzei

\usepackage{comment}

\theoremstyle{definition}

\newcommand{\ee}{\mathrm{e}}

\newcommand{\fs}{\,fs}
\newcommand{\eV}{\,eV}
\newcommand{\alat}{a_\text{lat}}

\newcommand{\hf}{H^\text{F}}

\newcommand{\bk}{\bm{k}}
\newcommand{\br}{\bm{r}}

\newcommand{\bj}{\bm{j}}
\newcommand{\bB}{\bm{B}}
\newcommand{\bR}{\bm{R}}

\newcommand{\bjnd}{\bm{j}_n^\text{C}}
\newcommand{\bjnc}{\bm{j}_n^\text{IC}}

\begin{document}
\title{
Disorder Effects on the Origin of High-Order Harmonic Generation in Solids}
\author{Koki Chinzei}
\affiliation{Institute for Solid State Physics, University of Tokyo, Kashiwa, Chiba 277-8581, Japan}

\author{Tatsuhiko N. Ikeda}
\affiliation{Institute for Solid State Physics, University of Tokyo, Kashiwa, Chiba 277-8581, Japan}
\date{\today}
\begin{abstract}
We consider noninteracting electrons coupled to laser fields,
and study perturbatively the effects of the lattice potential involving disorder
on the harmonic components of the electric current, which are sources of high-order harmonic generation (HHG).
By using the Floquet-Keldysh Green functions,
we show that each harmonic component consists of the coherent and the incoherent parts,
which arise respectively from the coherent and the incoherent scatterings by the local ion potentials.
As the disorder increases,
the coherent part decreases, the incoherent one increases,
and the total harmonic component of the current 
first decreases rapidly and then approaches a nonzero value.
Our results highlight the importance of the periodicity of crystals,
which builds up the Bloch states extending over the solid.
This is markedly different from the traditional HHG in atomic gases,
where the positions of individual atoms are irrelevant.

\end{abstract}
\maketitle

%##########################################################
%########  INTRODUCTION  ##################################
%##########################################################
%#### General introduction ###
\section{Introduction}
%#### General introduction ###
High-order harmonic generation (HHG),
which was traditionally studied in atomic and molecular gases~\cite{Brabec2000},
has recently been extended to solids~\cite{Ghimire2011,Schubert2014,Hohenleutner2015,Luu2015,Ndabashimiye2016,Vampa2015,You2017,Yoshikawa2017,Higuchi2017,Kaneshima2018}.
Owing to the state-of-the-art optics technology,
HHG is important not only as the foundation of the attosecond physics~\cite{Ghimire2014,Calegari2016}
but also as a probe to study electron dynamics in solids
under strong laser electric fields.
HHG from solids, unlike those from gases, 
should reflect band structures and crystal symmetries~\cite{Vampa2015a},
and active studies are ongoing to elucidate their principles and applications.

%#### finite# band models : merit simple pics. & experiments ###
Mechanisms and characteristics of HHG in solids
are often theoretically studied by effective two-band models
or, equivalently, tight-binding models.
For semiconductors,
two-band models~\cite{Golde2006,Golde2008,Golde2011,Vampa2014,Vampa2015c,Higuchi2014,Tamaya2016a,Tamaya2016b}
revealed that both the interband transitions and the intraband dynamics are important sources of HHG
although their relative importances seem intricate.
The intraband contribution is enhanced when the band dispersion has more anharmonicity~\cite{Ghimire2011},
whereas the interband contribution remains significant even without intraband dynamics in the valence band~\cite{Osika2017,Catoire2018}.
The mechanism of HHG in semiconductors is still under active debate.
The effective two-band models are also useful to study HHG in, for example,
graphene~\cite{Mikhailov2008,Mikhailov2016}, Mott insulators~\cite{Murakami2018b,Murakami2018a}, charge-density-wave materials~\cite{Nag2018,Ikeda2018a}, superconductors~\cite{Kawakami2018,Yonemitsu2018}, and topological insulators~\cite{Bauer2018,Jurss2019},
where both the inter- and the intra-band contributions play important roles.
We emphasize that both contributions originate from the periodic lattice potential,
which is presupposed and expressed by the band dispersions and the interband coupling in those effective models.

%#### infinite# band : merit principles
%#### However, why? hasn't been elucidated
To discuss the very origin of HHG in solids,
different kinds of models are used,
where the periodic lattice potential appears
explicitly in the Hamiltonian~\cite{Plaja1992,Faisal1997,Faisal2005,Yan2008,Korbman2013,Park2014,Wu2015,Du2017,Ikemachi2017,Jia2017,Ikemachi2018,Ikeda2018,Navarrete2019}.
In these models, it is manifest that there is no harmonic generation in the absence of the periodic lattice potential
no matter how strong the laser field becomes (see e.g. Ref.~\cite{Ikeda2018}).
Once the lattice potential is introduced,
the quadratic energy dispersion is folded to form the energy bands,
and the laser field causes both the anharmonic intraband dynamics and the interband transitions~\cite{Wu2015,Du2017,Ikemachi2017}.
HHG has been studied in these models
and the results seem consistent with those of the effective models.
However, it has not been well studied how HHG changes when the lattice potential is not perfectly periodic.
Given that the periodic lattice potential is the origin of HHG in solids,
how important is the perfect periodicity of the lattice potential?

In this paper, we address this question and investigate the origin of HHG in solids
by considering
the high-harmonic current (HHC),
which gives rise to HHG,
under strong laser fields on a weak lattice potential involving disorder.
By invoking the Floquet-Keldysh Green functions,
we show that the HHC consists of the coherent and incoherent parts,
and rapidly decreases to a nonzero value as the disorder increases.
On the basis of these results, we highlight the difference between  
the mechanisms of the HHG in solids and gases: the positions of atoms are relevant in solids while irrelevant in gases.
These results show the importance of the periodicity of the lattice potential in HHG from solids,
and reinforce the fact that its origin is the coherent dynamics of the Bloch state.

This paper is organized as follows.
In Sec. II, we formulate the problem of calculating HHG from disordered solids by using the Floquet-Keldysh formalism.
In Sec. III, we analyze the disorder effects on HHG by both analytical and numerical approaches,
uncovering the coherent and incoherent nature of HHG in solids.
In Sec. IV, we discuss the essential difference between HHG in solids and gases by using our results.
Finally in Sec.~\ref{sec:conclusions}, we summarize our study with concluding remarks.

%################################################################
%#####
\section{Formulation of the problem}
In this section, we formulate the problem of calculating HHC driven by a strong ac electric field in the presence of the disordered lattice potential. 
We begin by solving the problem in the absence of the potential, and show that there is no HHC.
Then we introduce our model of the lattice potential [see Eq.~\eqref{eq:potential}], 
and derive the formula for the HHC in terms of the Floquet-Keldysh Green function.

\subsection{Electron dynamics without potential}
We begin
by analyzing the electron dynamics in the absence of the lattice potential.
Let us consider noninteracting electrons coupled to a
homogeneous ac electric field at frequency $\Omega$
in $d$ dimensions.
We ignore the spin degree of freedom since it merely doubles our results.
We represent the ac field by
the vector potential $\bm{A}(t)=\bm{A}_0 \cos \Omega t$ in the velocity gauge.
The Hamiltonian is given by 
\begin{align}
\hat{H}_0(t)= \frac{\hat{\bm{p}}^2}{2m} -\frac{e}{m} \hat{\bm{p}} \cdot \bm{A}(t), \label{H0}
\end{align}
where $m$ and $e$ $(<0)$ are the mass and the electric charge of the electron, respectively,
and $\hat{\bm{p}}$ is the momentum operator.

%#####
The solutions of the time-dependent Schr\"{o}dinger equation,
$i\partial_t \psi(t) = \hat{H}_0(t) \psi(t)$, are known as the Volkov states~\cite{Keldysh1965,Faisal1973,Reiss1980}
($\hbar=1$ throughout this paper).
Their wave functions are characterized by the momentum $\bm{k}$ 
and given by
\begin{align}
\psi_{\bk}(\br,t) 
&= e^{-i \epsilon_k t} e^{i \bm{k} \cdot \bm{r}} \sum_n J_n(\alpha_{\bm{k}})e^{in\Omega t},\label{eq:volkov2}
\end{align}
where $\epsilon_{\bm{k}}=\bm{k}^2/2m$, $\alpha_{\bm{k}}=e \bm{A}_0 \cdot \bm{k}/(m\Omega)$, 
and $J_n(z)$ denotes the $n$-th Bessel function of the first kind.

%#####
The Volkov states carry no HHC
no matter how strong the ac electric field is.
In fact, their paramagnetic and diamagnetic currents
are given as
$\bm{j}_\text{para}(\bk,t)= -(ie/m) \int d\br \psi_{\bk}(\br,t)^\ast \nabla \psi_{\bk}(\br,t)=e\bk/m$
and $\bm{j}_\text{dia}(\bk,t) = -(e^2/m) \int d\br \psi_{\bk}(\br,t)^\ast \bm{A}(t) \psi_{\bk}(\br,t)=(e^2/m)\bm{A}(t)$
in appropriate wave-function normalizations.
These results explicitly show that the Volkov state carries no HHC,
or the Fourier components at $n\Omega$ with $|n|\ge 2$.

%##
We note that $\bm{j}_\text{dia}(\bk,t)=(e^2/m)\bm{A}(t)$
is satisfied not only by the Volkov state but also by any states.
Thus we ignore this part and focus on $\bm{j}_\text{para}(\bk,t)$ in the following.
Unlike $\bm{j}_\text{dia}(\bk,t)$, $\bm{j}_\text{para}(\bk,t)$
may involve harmonics when a lattice potential exists
and $\bm{k}$ is no longer a good quantum number.
Our problem is thus to study the Fourier components of
$\bm{j}_\text{para}(\bk,t)$ in the presence of the potential.
We remark that, when the lattice potential is periodic,
the energy bands are formed and
the paramagnetic current has both diagonal and off-diagonal
matrix elements in the band basis.
In other words,
the paramagnetic current involves 
both contributions from the interband transition and the intraband dynamics of electrons.

%#####
For later use,
we express the Volkov states in terms of the Floquet theory~\cite{Shirley1965}.
The Floquet Hamiltonian $\hf_0(\bk)$ corresponding to Eq.~\eqref{H0} reads
\begin{align}
\left[ \hf_0(\bk) \right]_{mn} 
&\equiv  m\Omega\delta_{mn} + \frac{\Omega}{2\pi} \int_0^{2\pi/\Omega} dt \ H_0(\bk,t) e^{i(m-n)\Omega t} \notag \\
&= (\epsilon_{\bk} + n\Omega)\delta_{mn} + \alpha_{\bk} (\delta_{m,n+1}+\delta_{m,n-1}),
\end{align}
where $m,n \in \mathbb{Z}$ are the Floquet indices, and  $H_0(\bk,t)$ is the Fourier component of $\hat{H}_0(t)$.
Its eigenvectors $\ket{\psi^M_0(\bk)}$ and eigenvalues $\epsilon^M_{\bk}$
are labeled by $M \in \mathbb{Z}$ and given by
\begin{align}
\ket{\psi^M_0(\bk)} &= \sum_n J_{M-n}(\alpha_{\bk}) \ket{\phi^n(\bk)}, \\
\epsilon^M_{\bk} &= \epsilon_k + M\Omega,\label{eq:quasi_volkov}
\end{align}
where $\ket{\phi^n(\bk)} $ is the Floquet state corresponding to the wave function $\propto e^{i\bk\cdot\br}e^{-i n\Omega t}$.
We note that $\ket{\psi^M_0(\bk)}$ is the Floquet representation of the Volkov state~\eqref{eq:volkov2}.

%%%%%%%%%%%%%%%%%%%%%%%%%%%%%%%%%%%%%%%%%%%%%%%%%%%%%%%%%%%%%%%%%%%%%%%%%%%%%%%%%%%%%%%%%%%%%%%%%%%%%%%%%%%%
\subsection{Model of lattice potential}
To investigate the effects of the lattice potential on the HHC,
we introduce the following lattice potential,
\begin{align}\label{eq:potential}
V(\br) &= U \sum_a  u(\bm{r}-\bm{r}_a),
\end{align}
where $U$ denotes the strength of the potential,
$u(\bm{r})$ is a dimensionless function localized at $\bm{r} \sim 0$,
and $\bm{r}_a$ denotes the position of the scattering center,
i.e. the ion in solids
\footnote{
We neglect the changes of the potential amplitude $U$ and the local ion potential $u(\br)$ by the disorder.
This simplification has also been made in theoretical studies of the solid-state HHG (see e.g. Ref.~\cite{Yu2019})
}.
Without loss of generality,
we assume $\int d\br\, u(\br) =0$
since a constant shift of the total energy changes no physical consequences.
This assumption implies that
the Fourier component $V_{\bk} = U \sum_a u_{\bk} e^{-i \bk \cdot \br_a}$
vanishes at $\bk=0$,
and thus the diagonal matrix elements of the lattice potential vanish in the momentum basis: $\braket{\bk | \hat{V} | \bk}=V_{\bk-\bk}=0$, where $\ket{\bk}$ is the momentum eigenstate with eigenvalue $\bk$.

\begin{figure}[t]
\center
\includegraphics[width=8.6cm]{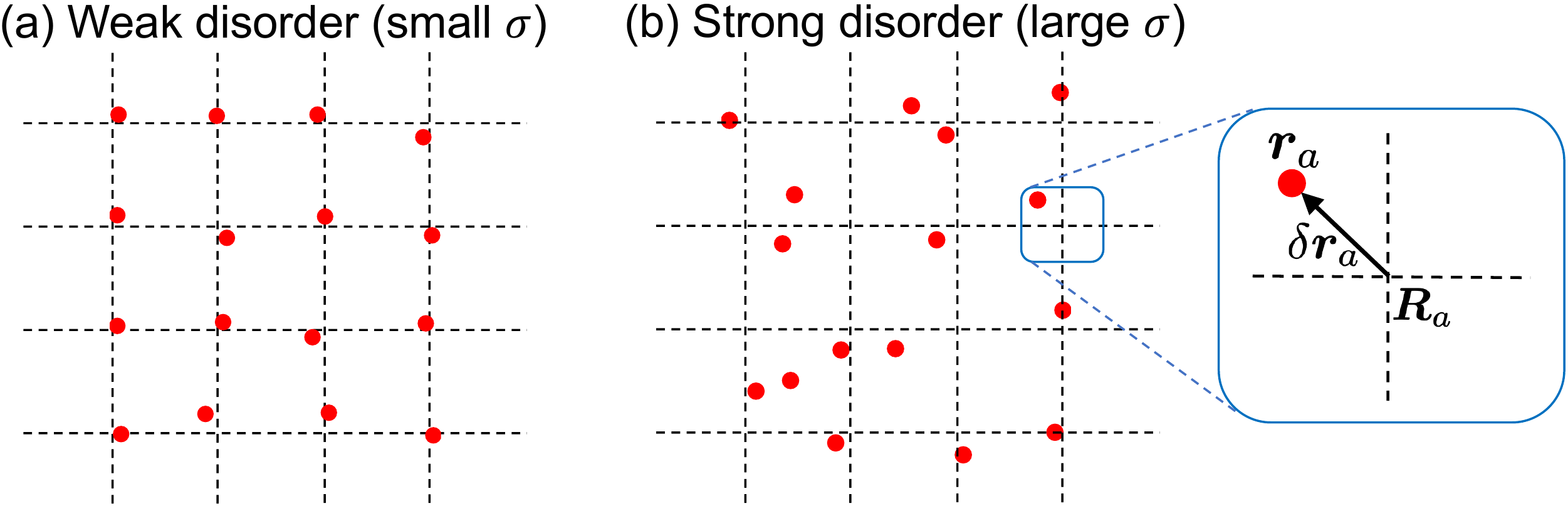}
\caption{
Schematic illustration of disordered lattice potential [Eq.~\eqref{eq:potential}] for square lattice in two dimensions.
Red dots indicate the positions of the local ion potentials for the weak disorder (a) and the strong disorder (b).
The inset shows our parametrization of each position.
}
\label{fig:disorder}
\end{figure}

%###### Paragraph to explain our potential #######
%##### Bravais lattice + fluctuation ######
Suppose that $\bm{r}_a$'s form an approximate Bravais lattice
such as the simple square lattice (see Fig.~\ref{fig:disorder}).
Let $\bm{R}_a$ denote the position
of each lattice point $a$ of the Bravais lattice,
which
is characterized by a set of $d$ integers $\{c_a^i \in \mathbb{Z} \mid i=1, 2, \cdots ,d \}$
as $\bm{R}_a = \sum_{i=1}^d c_a^i \bm{a}_i$
with the primitive vectors $\bm{a}_i$.
Then we introduce a small deviation $\delta\bm{r}_a$
and define $\br_a$ as $\bm{r}_a = \bm{R}_a + \delta \bm{r}_a$.
Unlike $\bm{R}_a$'s,
$\br_a$'s do not have the exact discrete translational symmetry.
%追加
We note that our model describes the amorphous solids where the ions do not array periodically rather than the doped semiconductors~\cite{Huang2017,Almalki2018,Yu2019} where the impurity potentials are periodically added and the electrons are trapped to them.

%#####
We assume that each of $\delta \bm{r}_a$'s
is an independent Gaussian random variable.
Its probability density function is given by
$P(\delta \bm{r}_a) =(2\pi\sigma^2)^{-1/2} \exp\left[- \delta r_a^2/(2\sigma^2) \right]$,
where the standard deviation $\sigma$
quantifies the randomness of the lattice.
Our assumption of the independence of the variables
means, for instance,
$\langle \delta \br_a \delta\br_b \rangle =\langle \delta \br_a \rangle \langle \delta \br_b \rangle $ for $a\neq b$,
where $\langle \cdots \rangle$ denotes the average
over the random variables.
In the following, we analyze the HHC
for a given set of $\{\br_a\}_a$
and take its average over the deviations $\{\delta \bm{r}_a\}_a$.

\begin{figure}[t]
\center
\includegraphics[width=8cm]{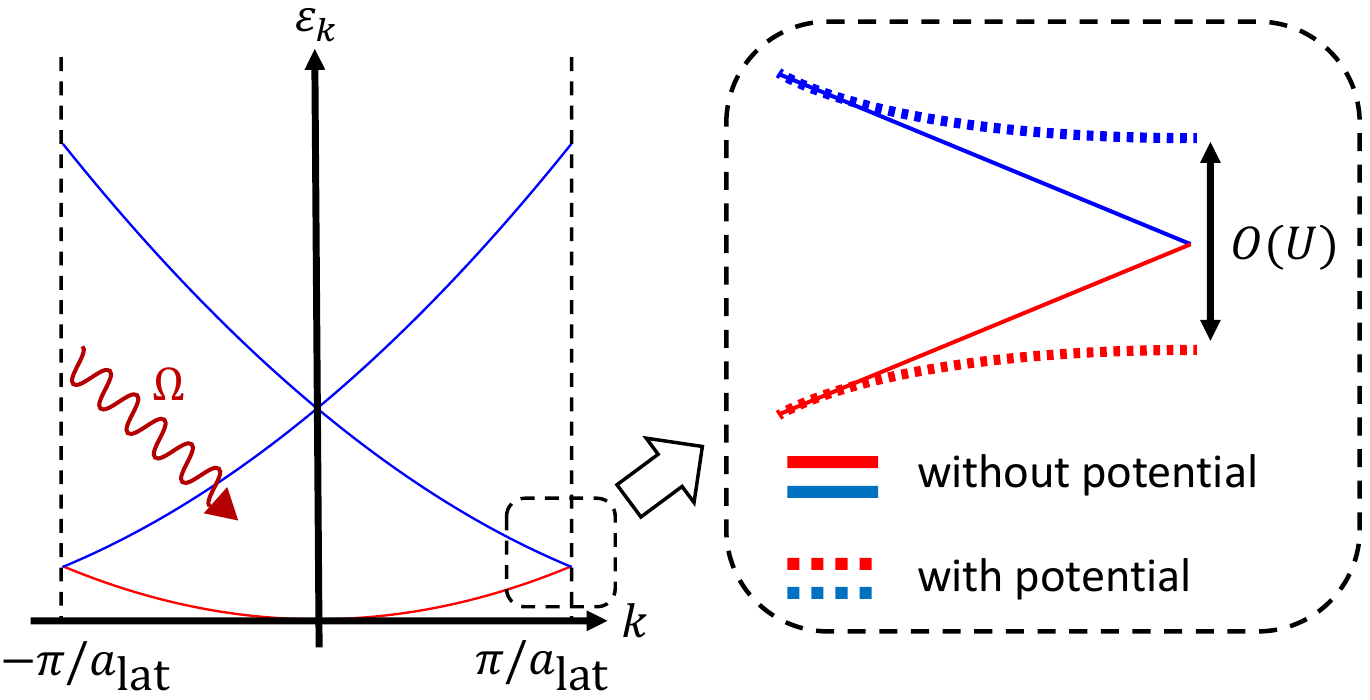}
\caption{
Folded parabolic energy dispersion in one dimension. The inset shows an energy gap opened by the lattice potential without disorder.
}
\label{fig:bandstructure}
\end{figure}

Here we show the band structure of our model.
In the absence of the lattice potential $V(\br)$, the energy dispersion is parabolic as shown in Fig.~\ref{fig:bandstructure}, where the parabolic dispersion is folded into the first Brillouin zone for convenience.
In the presence of $V(\br)$ without disorder (when there is disorder, the band picture is no longer valid since the discrete translational symmetry is broken),
the energy gaps open at the degeneracy points of the folded dispersion, and the energy bands are formed as schematically shown in the inset of Fig.~\ref{fig:bandstructure}. 
We note that the energy gap is $O(U)$ when the lattice potential is a perturbation as we assume later.
This means that our model is closer to narrow-gap semiconductors, metals, or Dirac materials rather than semiconductors.

%#### 
%#####

%#### 
%#####

%##########################################
\subsection{Floquet-Keldysh formalism}
We analyze the HHC 
by using the Floquet-Keldysh formalism,
which is a method of combining the Floquet theory and the non-equilibrium Green function (see e.g., Refs.~\cite{Oka2009,Morimoto2016}).
The non-equilibrium Green function has three components, the retarded, the advanced and the Keldysh Green functions.
The retarded and the advanced Green functions have information of the energy spectrum,
whereas the Keldysh Green function has that of the energy spectrum and the distribution function.

Without the lattice potential,
the analytical solutions of the Green functions denoted by $g$ are available
since all the eigenstates of the Floquet Hamiltonian $\hf_0(\bk)$ are obtained.
The retarded and advanced Green functions are given by~\cite{Faisal1989,Faisal1991} (see also Ref.~\cite{Ikeda2018})
\begin{align}
g_{mn}^{R/A}(\bk,\omega) 
&= \sum_M \frac{ J_{M-m}(\alpha_{\bk}) J_{M-n}(\alpha_{\bk}) }{\omega - \epsilon_{\bk}^M \pm i \eta}, \label{eq:gR}
\end{align}
where $m, n \in \mathbb{Z}$ are the Floquet indices.
We note that the band degrees of freedom are included in $\bk$; for instance, $|\bk| < \pi/a_\text{lat}$ ($\pi/a_\text{lat} < |\bk| < 2\pi/a_\text{lat}$) corresponds to the lowest band (the second-lowest band) in Fig.~\ref{fig:bandstructure}.
Here, we assume a finite relaxation time $1/\eta$ $(>0)$, which stems from the electron correlation and so on.
Because of the finite relaxation time, the Keldysh component has a Lorentzian form,
\begin{align}
g_{mn}^{K}(\bk,\omega) 
&= -2 i \eta (1-2n_{\bk}) \sum_M  \frac{J_{M-m}(\alpha_{\bk}) J_{M-n}(\alpha_{\bk}) }{(\omega - \epsilon_{\bk}^M)^2 +  \eta^2}, \label{eq:gK}
\end{align}
where $n_{\bk}$ is a distribution function.
In Sec.~\ref{sec:numerics} of the numerical calculation part, we assume that only the lowest band is occupied by the electrons and the other bands are empty (see Fig.~\ref{fig:bandstructure}): $n_{\bk}=1$ for $|\bk|=k \leq k_F$ and $n_{\bk}=0$ for $k > k_F$,
where $k_F = \pi/a_\mathrm{lat}$ is the Fermi momentum and $a_\mathrm{lat}$ is the lattice constant.

Once we have the non-equilibrium Green functions,
the Fourier components of the paramagnetic current $\bj_\text{para}(\bk,t)$ at frequency $n\Omega$ is obtained as
\begin{align}
\bj_n(\bk) = -i \int_{-\infty}^{\infty} \frac{d\omega}{2\pi} v_{\bk} g^<_{n0}(\bk,\omega), \label{eq:jn}
\end{align}
where $v_{\bk}=e\bk/m$ is the electron velocity,
and $g^<=(g^A-g^R+g^K)/2$ is the lesser Green function [See Appendix A for the derivation of Eq.~\eqref{eq:jn}].
From Eq.~\eqref{eq:gR}-\eqref{eq:jn}, we have $\bj_n(\bk)=\delta_{n0} n_{\bk} e\bk/m$,
which is consistent with $\bm{j}_\text{para}(\bk,t)=e\bm{k}/m$ obtained above
directly by the wave function~\eqref{eq:volkov2}.
We note that, in the absence of the lattice potential, the net paramagnetic current $\bj_n = \int d^d k/(2\pi)^d \bj_n(\bk)$ vanishes because $\bj_n(\bk) = -\bj_n(-\bk)$
if $n_{\bk}=n_{-\bk}$
(Note that the diamagnetic current $\bj_\text{dia}$
exists but involves only the fundamental frequency $\Omega$
as noted above).

%##########################################################################
%##########################################################################
%#### 

%#### 
\begin{figure}[t]
\center
\includegraphics[width=8.6cm]{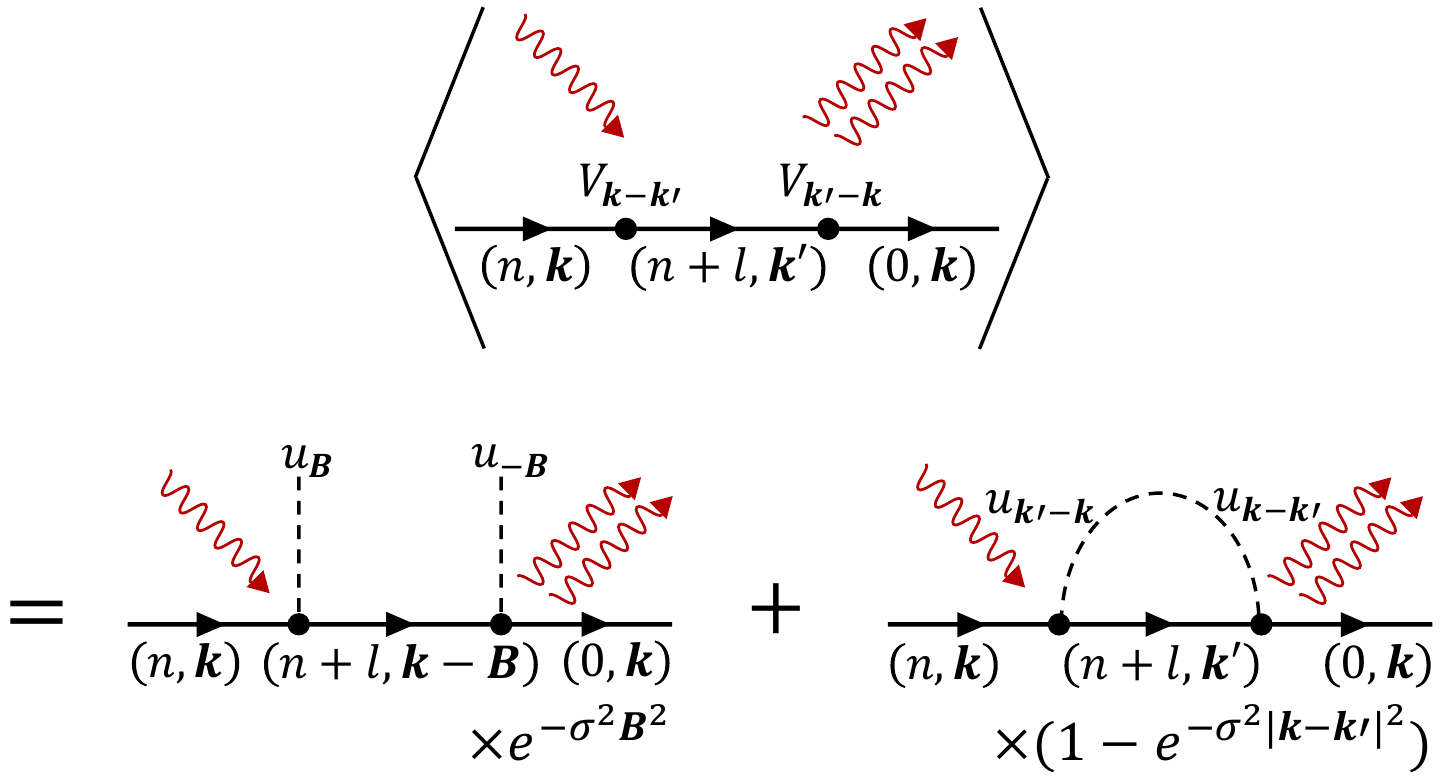}
\caption{
Feynmann diagram for $O(U^2)$ contribution to $\bj_n(\bk)$.
The solid line with $(n,\bk)$ indicates the $n$-photon state with momentum $\bk$,
and the dot denotes the potential scattering.
%We sum up the photon number $n+l$ and the momentum $\bk'$ in the internal line.
After random averaging,
$\langle  \bj_n(\bk)\rangle$
consists of the two diagrams [see Eqs.~\eqref{eq:jnbar}-\eqref{eq:fn}].
The first (second) diagram on the right-hand side corresponds to the two scatterings
occurring at different sites (the same site).
}
\label{fig:diagram}
\end{figure}
%#####

%#########################################
\section{Analysis of high-order harmonic currents}
In this section, we analyze the HHC in the presence of the disordered lattice potential on the basis of the Floquet-Keldysh formalism. 
We perform the leading-order calculation of
the Green functions and derive the expressions for the
HHC [see Eqs.~\eqref{eq:jnbar}--\eqref{eq:fn}]. 
We then numerically evaluate
those expressions and discuss the disorder effects on the
HHC.

\subsection{Analytical calculations of HHC}
In the presence of the lattice potential~\eqref{eq:potential},
it is difficult to obtain the exact solutions to the Green functions,
which are denoted by $G$.
We invoke the perturbation theory in terms of the potential amplitude $U$
to approximately obtain $G$,
and calculate the leading-order correction of the HHC induced by the potential
by Eq.~\eqref{eq:jn} with $g^<$ replaced by $G^<$.

%########
The HHC does not appear up to the first order of $U$.
In fact, at the zeroth order, $G$ corresponds to $g$
and there is no HHC as mentioned above.
Besides, the $O(U)$ corrections to the Green functions
induce no HHC.
This is because
$V_{\bk}$ vanishes at $\bk=0$
and
the corrections are made only on
the off-diagonal components $G_{\bk \bk'}$ ($\bk\neq\bk'$),
while
the HHC originates from the diagonal elements [see Eq.~\eqref{eq:jn}].

%####
The HHC first appears at the second order.
According to the second-order Feynmann rules for the non-equilibrium Green function~\cite{Rammer1986},
the $O(U^2)$ corrections to the full Green functions are given by
\begin{align}
&G^{(2)}_{mn} (\bk,\omega) \notag \\
&= \sum_{p,q} \int \frac{d^d k'}{(2\pi)^d} g_{mp}(\bk,\omega) V_{\bk-\bk'} g_{pq}(\bk',\omega) V_{\bk'-\bk} g_{qn}(\bk,\omega), \label{eq:G2pert}
\end{align}
where both $G^{(2)}_{mn}(k,\omega)$ and $g_{mn}(k,\omega)$ are the $2\times 2$ matrices,
\begin{align}
G_{mn}^{(2)} (\bk,\omega) 
&= 
\begin{pmatrix}
G^R & G^K \\
0 & G^A
\end{pmatrix}_{mn}^{(2)}  (\bk,\omega), \\ 
g_{mn} (\bk,\omega) 
&= 
\begin{pmatrix}
g^R & g^K \\
0 & g^A
\end{pmatrix}_{mn} (\bk,\omega).
\end{align}
The technical details to derive and evaluate Eq.~\eqref{eq:G2pert} are shown in Appendix~\ref{sec:calcGF}.

To obtain the results for the HHC, we take the average of the HHC over the random variables $\delta \br_a$'s,
\begin{align}\label{eq:jnavg}
\braket{\bj_n} = -i \int \frac{d^d k}{(2\pi)^d} \int \frac{d\omega}{2\pi} v_{\bk} \braket{ G^{<(2)}_{n0}(\bk,\omega) }.
\end{align}
We note that the random variables appear only in the scattering vertices $|V_{\bk-\bk'}|^2$ in Eq.~\eqref{eq:jnavg} in the following form: 
\begin{align}
&\sum_{a,b} \braket{e^{-i(\bk-\bk') \cdot \br_a} e^{i (\bk-\bk') \cdot \br_b}} \notag \\
&= \sum_{\bB} \delta(\bk - \bk' +\bB) e^{-\sigma^2 |\bB|^2} + (1 - e^{-\sigma^2 |\bk-\bk'|^2}),\label{eq:phase_avg}
\end{align}
where $\bB$ denotes the reciprocal vector.
To obtain Eq.~\eqref{eq:phase_avg}, we have used $\braket{e^{i\bk \cdot \delta \br_a}} = e^{-\sigma^2 \bk^2/2}$ and $\sum_a e^{i\bk \cdot \bR_a} = \sum_{\bB} \delta(\bk+\bB)$.
The first and the second terms on the right-hand side of Eq.~\eqref{eq:phase_avg} correspond respectively to the terms with $a\neq b$ and $a= b$ on the left-hand side.
In other words,
these terms imply the double potential scattering at two different sites and the same site.

Corresponding to this decomposition, we decompose the HHC into two parts, 
\begin{align}
\langle\bj_n\rangle = \bjnd + \bjnc, \label{eq:jnbar}
\end{align}
where $\bjnd$ ($\bjnc$) stems from the double scattering at two different sites (the same site),
and has the coherent (incoherent) nature as discussed below
\footnote{
The coherent (incoherent) current $\bjnd$ ($\bjnc$) is contributed by both the intra- and the inter-band dynamics in the three-step model in the momentum space~\cite{Vampa2014}.
In the real space, the HHG is produced by the electrons recolliding with the various ion sites.
As discussed in Ref.~\cite{Ghimire2019},
these two interpretations describe the same phenomenon and thus are consistent with each other
(see also Discussions section in the main text).
}. 
Each current is given by
\begin{align}
\bjnd &=  \int \frac{d^d k}{(2\pi)^d}  \sum_{\bB} v_{\bk} f_n(\bk,\bk-\bB) e^{-\sigma^2 \bB^2}, \label{eq:jnd} \\
\bjnc &=  \int \frac{d^d k}{(2\pi)^d} \int \frac{d^d k'}{(2\pi)^d} v_{\bk} f_n(\bk,\bk') (1-e^{-\sigma^2 |\bk-\bk'|^2}), \label{eq:jnc} 
\end{align}
where $f_n(\bk,\bk')$ is 
\begin{widetext}
\begin{align}
f_n(\bk,\bk') 
&=  U^2 |u_{\bk-\bk'}|^2 (n_{\bk} - n_{\bk'}) \left( \frac{n\Omega - 4i\eta}{n\Omega - 2i\eta} \right)
\sum_l \frac{ J_{-l}(\alpha_{\bk-\bk'}) J_{n+l}(\alpha_{\bk'-\bk})}{ \left[(\epsilon_{\bk} - \epsilon_{\bk'}) - l\Omega - 2i\eta \right] \left[(\epsilon_{\bk'} - \epsilon_{\bk}) + (n+l)\Omega - 2i\eta \right] }. \label{eq:fn}
\end{align}
\end{widetext}
These relations are the main results
of the present work.
We note $\langle \bj_n\rangle=0$ for even $n$'s when the system is inversion symmetric after random averaging.
One can prove this by noticing that contributions from $\pm \bk$ cancel out each other.

%####
We remark the similarity and the difference between $\bjnd$ and $\bjnc$.
The similarity is that the $n$-th component derives from the Feynman diagram
in which the initial and the final momenta are the same but the Floquet indices (``photon numbers'')
differ by $n$ (see Fig.~\ref{fig:diagram}).
The difference is that $\bjnc$ is contributed by any internal momentum
whereas $\bjnd$ by the discrete internal momenta shifted by reciprocal lattice vectors.
This difference reflects the coherent (incoherent) nature of the potential scatterings
in $\bjnd$ ($\bjnc$).
The physical meaning of the scattering becomes more obvious by introducing the band picture (see Fig.~\ref{fig:bandstructure}).
The scattering process in $\bjnd$ with the reciprocal lattice vector $\bB$ can be interpreted as the interband transition without changing the lattice momentum, or $\bk$ modulo $\bB$.
%This invariance of the lattice momentum implies the recovery of the periodicity after random averaging.
On the other hand, the incoherent current $\bjnc$ is contributed by the scattering with any $\bk'$ and thus involves both the inter- and intra-band transitions.

%TODO: combine and shorten
%###########
When the disorder is very small, or $\sigma\simeq0$,
the $n$-th harmonic current $\langle \bj_n\rangle$
is dominated by $\bjnd$.
In the limit of $\sigma\to0$,
$\bjnc$ vanishes and $\bjnd$ reduces to the result
for the perfectly periodic lattice.
In this limit, $\bjnd$ is the largest because
the phase factor $e^{i(\bk-\bk')\cdot (\br_a-\br_b)}$ 
for the double scattering at any pair of sites $a$ and $b$ becomes unity.
Namely, the scatterings at different sites are all coherent.
As the disorder increases,
$\bjnd$ exponentially decays in $\sigma^2$.
This is because 
the vertex phase factors $e^{i(\bk-\bk')\cdot \br_a}$ fluctuate
and the scatterings at different pairs of sites work destructively.

%######
On the other hand,
$\bjnc$ becomes dominant when the disorder is very large.
In the limit of $\sigma \rightarrow \infty$,
$\bjnd$ vanishes
and $\bjnc$ converges to a nonzero value.
In this limit, the fluctuations $\delta \br_a$ are so large
that the phase factor $e^{i(\bk-\bk')\cdot (\br_a-\br_b)}$
is nonvanishing only for $a=b$.
In other words, $\bjnc$ consists of the incoherent sum
of the contributions from each scattering center.
This local nature of $\bjnc$ manifests as
the presence of any momentum $\bk'$ in Eq.~\eqref{eq:jnc}.
We note that $\bjnc$ vanishes at $\sigma=0$,
where the lattice potential is perfectly periodic.
Thus $\bjnc$ is specific to disordered systems.

%######
\begin{figure}[t]
\center
\includegraphics[width=8.6cm]{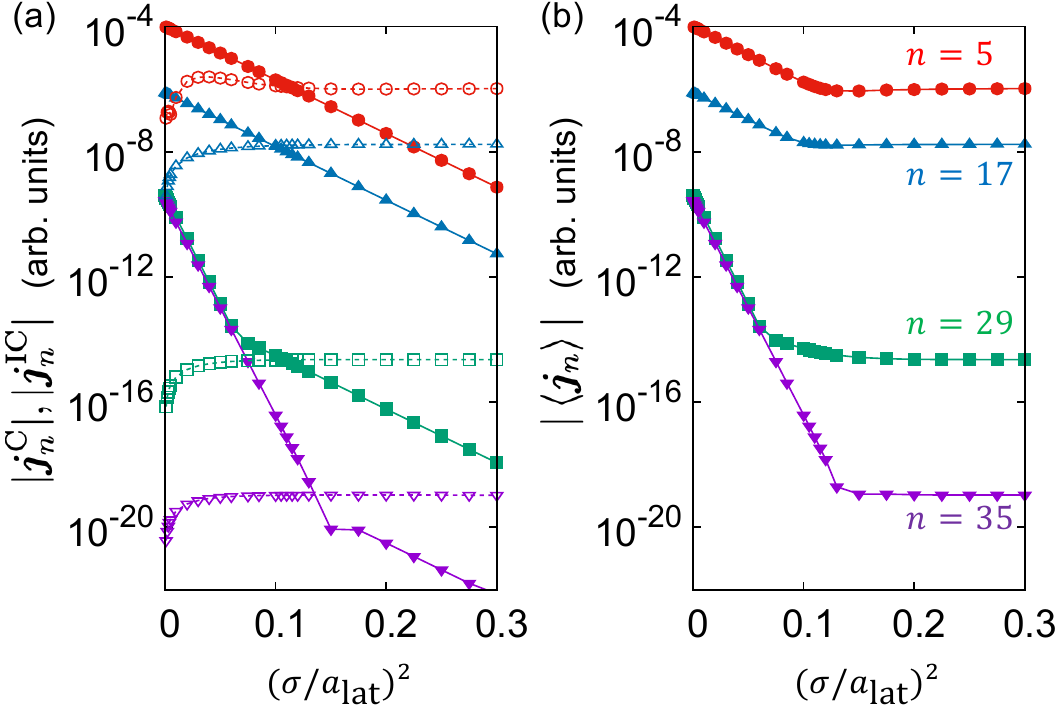}
\caption{
(a) Disorder dependence of coherent part $|\bjnd|$ (filled)
and incoherent part $|\bjnc|$ (open)
calculated by Eqs.~\eqref{eq:jnd} and \eqref{eq:jnc}
for a model local potential $u(x)$ in one dimension.
(b) Total HHC calculated by Eq.~\eqref{eq:jnbar}.
In both panels, we set $E_0 =10.6\,\mathrm{MV/cm}$
and the harmonic orders are
$n=$5 (circle), 17 (triangle), 29 (square), and 35 (inverted triangle).
}
\label{fig:sigma}
\end{figure}
%#####

%######
\begin{figure}[t]
\center
\includegraphics[width=8cm]{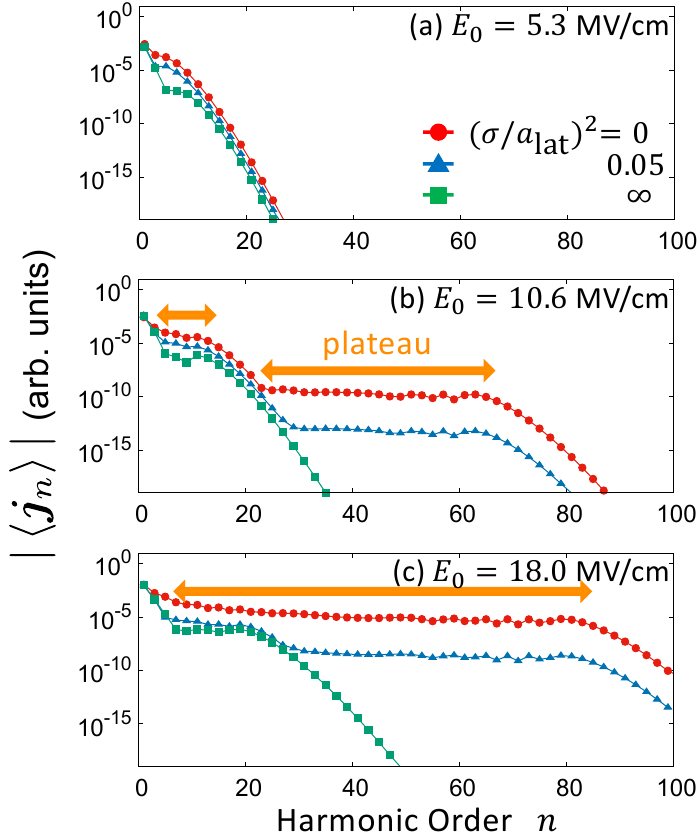}
\caption{
Harmonic spectrum for $(\sigma/a_\text{lat})^2 = 0$ (circle), 0.05 (triangle), and $+\infty$ (square).
Each panel corresponds to the case of (a) $E_0 = 5.3 \,\mathrm{MV/cm}$, (b) $E_0 = 10.6 \,\mathrm{MV/cm}$, and (c) $E_0 = 18.0 \,\mathrm{MV/cm}$.
}
\label{fig:N}
\end{figure}
%#####

%#########################################
\subsection{Numerical evaluation and quantitative analysis}\label{sec:numerics}
Now
we numerically evaluate Eqs.~\eqref{eq:jnd} and \eqref{eq:jnc}
for a choice of the potential $u(\br)$ and perform quantitative analyses.
We work in one dimension for simplicity.
We adopt $u(x) = e^{-32x^2/a_\text{lat}^2}\cos(16x/a_\text{lat}) + e^{-32x^2/a_\text{lat}^2-2}$,
which is localized around $x=0$ and satisfies $\int dx\,u(x)=0$.
Here $a_\text{lat}=5\,\text{\AA}$ is a typical lattice constant
and we set the parameters as $k_F/\alat=\pi$, $\hbar\Omega=0.27$\eV, and $\tau=1/\eta=48$\fs~\cite{Golde2008}.
Since $\hbar\Omega$ is above the band gap of $O(U)$,
our setup is closer to narrow-gap semiconductors, metals, or Dirac materials rather than semiconductors.
Thus, in applying our ideal model calculations
directly to experimental situations,
one should be careful
because a very intense laser may cause the damage of sample solids,
which is neglected in our calculation.

%###########
The crossover between the two limits $\sigma=0$ and $\sigma\to\infty$
is shown in Fig.~\ref{fig:sigma}(a), 
where $|\bjnd|$ and $|\bjnc|$ evaluated by Eqs.~\eqref{eq:jnbar}--\eqref{eq:fn} are plotted against $\sigma$ for $n=5, 17, 29$, and $35$.
As discussed above,
$|\bjnd|$ exponentially decays
whereas $|\bjnc|$ firstly increases at $\sigma/a_\text{lat} \sim 0$ and approaches a nonzero value.
The coherent part is larger than the incoherent one for $(\sigma/a_\text{lat})^2 \lesssim 0.1$,
and vice versa for $(\sigma/a_\text{lat})^2 \gtrsim 0.1$.
This leads to the crossover of the HHC between the regimes where the coherent and the incoherent parts are dominant.

The harmonic current $|\langle \bj_n\rangle|$, which is the sum of $\bjnd$ and $\bjnc$, is plotted against $\sigma$ for $n=5, 17, 29$, and $35$ in Fig.~\ref{fig:sigma}(b).
From the figure, we confirm that 
$|\langle \bj_n\rangle|$ decreases exponentially at $\sigma \sim 0$ (see also Ref.~\cite{Orlando2018}),
and approaches a nonzero value as $\sigma$ increases.
This is the crossover from the coherent scattering regime to the incoherent one.
Also, we emphasize that the harmonic currents $\langle \bj_n\rangle$ at $\sigma = 0$ is much larger than those at $\sigma \rightarrow \infty$.
This result shows the importance of the periodicity of the lattice potential to obtain large HHG.

We remark that the initial decrease of the HHC is more rapid for the higher order $n$.
In fact, at small $\sigma$ in Fig.~\ref{fig:sigma}(b), 
we observe $|\langle \bj_n\rangle| \propto \ee^{-(2\pi\sigma/\alat)^2}$ for $n=5$ and 17
whereas $|\langle \bj_n\rangle| \propto \ee^{-(4\pi\sigma/\alat)^2}$ for $n=29$ and $35$.
This implies that the low-(high-)order coherent current is dominantly contributed from the potential scattering
with $B=2\pi/a_\text{lat}$ $(4\pi/a_\text{lat})$ in Eq.~\eqref{eq:jnd}.
In terms of the band picture, the scattering with $B=2\pi/a_\text{lat}$ ($4\pi/a_\text{lat}$) corresponds to the scattering from the lowest band to the second-lowest band (the third-lowest band).

%#########################################
The disorder has nontrivial effects on the structure of the harmonic spectrum through this difference.
In Fig.~\ref{fig:N}(b) ($E_0 =A_0\Omega=10.6$\,MV/cm), we observe two plateaus in the absence of disorder $\sigma=0$.
As $\sigma$ increases,
the second plateau decreases more rapidly than the first one,
and the only first plateau remains in the limit of $\sigma\to\infty$, where $\bj_n=\bjnc$.
This difference between the decay speeds of the two plateaus is consistent with the observation in Fig.~\ref{fig:sigma}(b),
which shows that the higher-order harmonic currents decay more rapidly than the lower-order ones.
Therefore, as discussed above, these imply that the second plateau is made up by the coherent current due to $B=4\pi/\alat$
whereas the first plateau is by both the coherent part due to $B=2\pi/\alat$ and the incoherent one.

We show the results for a weaker or a stronger laser field.
Figure \ref{fig:N}(a) shows the result for a weaker laser field ($E_0 =5.3$\,MV/cm), 
where the second plateau vanishes even for $\sigma=0$.
This implies that the scattering with large $B$ is hard to occur for a weak laser because it needs the multiphoton absorption. 
For a stronger laser field ($E_0 =18.0$\,MV/cm), we obtain qualitatively similar result to that for $E_0 =10.6$\,MV/cm.
As shown in Fig.~\ref{fig:N}(c), the first and the second plateaus merge to form a wide plateau at $\sigma=0$.
However, the second plateau decreases to vanish as $\sigma\to\infty$,
and only the first plateau remains in the limit.
We note that the first plateau is wider for the stronger field $E_0$.

%#########################################
\section{Discussions}

%#### 
\begin{figure}[t]
\center
\includegraphics[width=8.6cm]{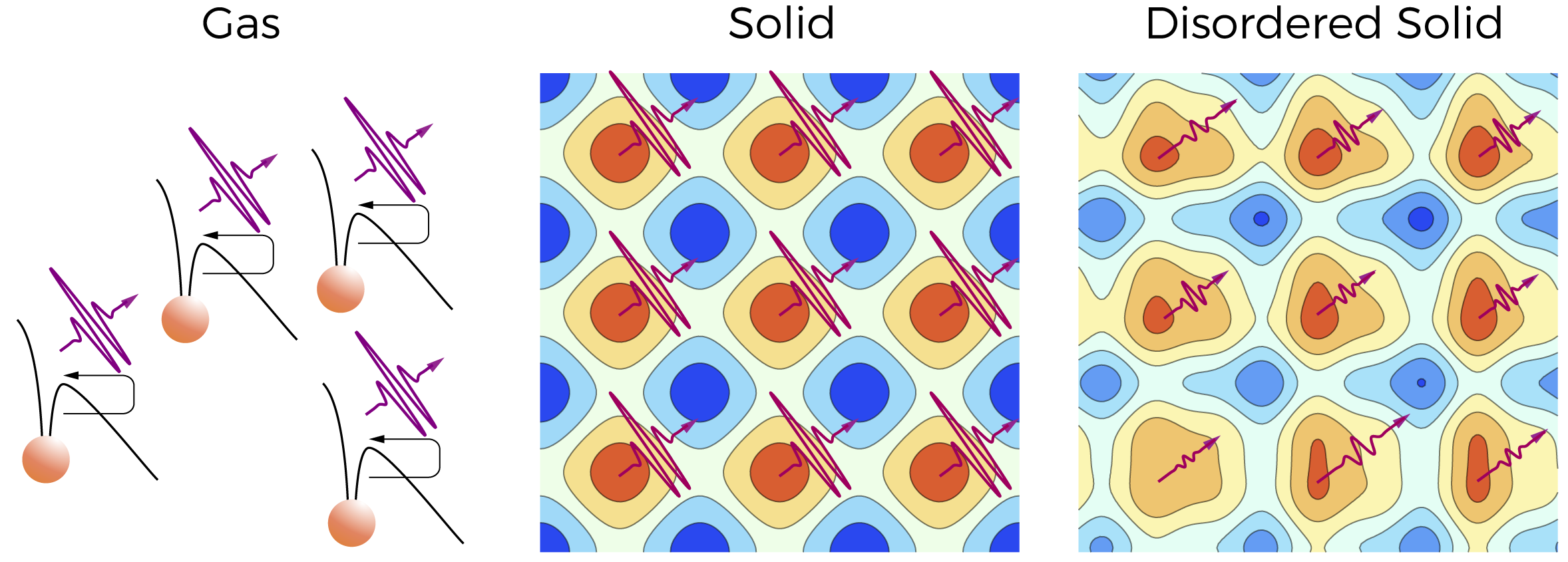}
\caption{
Schematic illustrations of mechanisms of HHG
in gases (a), solids (b),
and disordered solids (c).
(a) Each sphere shows an atom,
the thick arrow on it does its velocity,
the solid curves show the potential energy for the electron,
and the thin arrow does the electron trajectory.
(b) and (c)
The pattern illustrates the real-space distribution
of the wave function of the Floquet eigenstate
in the presence of the lattice potential.
}
\label{fig:hhg}
\end{figure}
%#####

Although our setup is different from that for semiconductors as mentioned above,
it is intriguing to point out similarities of our results with the HHG experiments on amorphous silica~\cite{You2017a,Jurgens2019}.
In this experiment~\cite{You2017a}, they have observed high harmonics with order $n < 34$ by the peak field of $E_0\sim200$\,MV/cm,
while their efficiency is lower than those in crystalline quartz. 
Our results have two features observed in this experiment.
First, we have shown that, as increases, the HHC decreases
but approaches a nonzero value rather than vanishes.
Second, our first plateau extends over $n>20$ at $E_0>18.0$\,MV/cm even in the presence of the disorder.
These two characteristics are in line with the surprising observation that
such a high order has been detected in amorphous solids.

Let us finally discuss
the difference between the mechanisms of the HHG in gases and solids.
The HHG in an atomic gas is explained
by the celebrated three-step model~\cite{Corkum1993}.
According to this theory,
the tunneling ionization occurs at an atom,
an electron propagates in the laser field,
and the electron and the ion recombine to produce radiation.
These processes occur at individual atoms
and their positions are irrelevant
as shown in Fig.~\ref{fig:hhg}(a).

%######
Our results highlight the essential difference 
of the HHG in solids from that in gases
\footnote{
In the limit of $\sigma \rightarrow \infty$,
our model corresponds to strongly disordered solids rather than gases.
This is because the HHG in our model derives from multiscattering processes at several ion sites
while that in gases does from individual atoms.
}.
In solids, the HHG is induced by the potential scatterings
by the ions, whose positions are of crucial importance.
The HHG becomes the largest when they are aligned periodically
and the scatterings at different positions are coherent.
To put this in the momentum space,
the electrons in the periodic potential are in the Bloch states
whose wave functions extend over the entire crystal and are compatible with the periodicity~\cite{Ghimire2019}.
When driven by a strong laser field,
the Bloch states produce the HHC
and hence the HHG as schematically depicted in Fig.~\ref{fig:hhg}(b).
When the scattering centers fluctuate and the disorder sets in, the HHC rapidly decreases.
As schematically illustrated in Fig.~\ref{fig:hhg}(c),
the disorder disturbs the coherence of the Bloch states,
and the resultant HHC, or the HHG, is suppressed.

%#########################################
\section{Conclusions}\label{sec:conclusions}
We have studied the disorder effects on the HHG, or the HHC, in solids
by considering the lattice potential~\eqref{eq:potential},
where the scattering centers
fluctuate around a Bravais lattice.
We have noted that,
if there is no potential,
the electrons are in the Volkov states
and the HHC does not exist
no matter how strong the laser field is.
Then we have turned on a weak potential
and analyzed the induced HHC 
by means of the perturbation theory.
In other words,
we have focused on the origin of the HHG in solids.

Our main results~\eqref{eq:jnbar}--\eqref{eq:fn}
have been obtained by the Floquet-Keldysh formalism,
stating that
the $n$-th harmonic current
$\langle \bj_n\rangle$
averaged over the fluctuations of the potential
is the sum of the coherent part $\bjnd$
and the incoherent one $\bjnc$.
Here
$\bjnd$ and $\bjnc$ are due to the double scattering at two different sites $a\neq b$ and the same site $a=b$, respectively.
In the absence of the disorder ($\sigma=0$),
where the lattice is perfectly periodic,
$\bjnd$ is the largest and $\bjnc$ vanishes.
As $\sigma$ increases,
$\bjnd$ decreases exponentially
and $\bjnc$ increases to be dominant.
Namely, the HHC exhibits a crossover
from the coherent to the incoherent ones
as the disorder increases.
We have numerically shown that the total HHC
$\langle \bj_n\rangle$
is the largest in the limit of no disorder.
Besides, we have shown that the disorder significantly deforms the structure of the harmonic spectrum.
As shown in Fig.~\ref{fig:N}, the first plateau is robust against the disorder whereas the second plateau is strongly suppressed by the disorder.

%#####
Our results
shed light on the difference
of the mechanisms of the HHG in gases and solids.
In atomic gases, the HHG occurs at individual atoms and their positions are not relevant.
In solids, however,
the positions of the scattering centers
are crucially important,
and the HHG is the largest when they
form a periodic lattice.
This implies
in the momentum space
that the periodic potential makes up the Bloch states with coherence,
which produce the largest radiation,
and the disorder breaks the coherence
to suppress the HHG.

Finally, we comment on the applicability of our results to experiments.
In this study, we have considered a simplified model of the disordered lattice potential
and analyzed the situation in which the photon energy is larger than the gap.
This situation is different from those of typical HHG experiments in semiconductors
and rather close to those in (semi)metals and narrow-gap semiconductors.
Thus one should be careful to compare the experimental results with our theoretical ones.
To study real disordered materials, one may need a more detailed Hamiltonian and possibly the interaction between electrons~\cite{Yu2019a}.
We leave this as an important open issue.

\section*{acknowledgements}
This work was supported by JSPS KAKENHI Grant No.~JP18K13495.
K. C. acknowledges financial support provided
by the Advanced Leading Graduate Course for Photon Science at the University of Tokyo.

%########################################################################
%########################################################################
%########################################################################
%########################################################################
%########################################################################
%\begin{widetext}
\appendix
%########################################################################
\section{Derivation of formula for HHC}
We derive the formula for the HHC described by the non-equilibrium Green function (see Eq.~(4) in the main text).
Let $\hat{\psi}(\br,t)$ denote the second-quantized field operator in the Heisenberg picture,
and then the electric current is given by
\begin{align}
&\bj(t) \notag\\
&= -\frac{ie}{2m} \int d^dr \left\langle \hat{\psi}^\dag(\br,t) [\nabla  \hat{\psi}(\br,t)] - [\nabla  \hat{\psi}^\dag(\br,t)] \hat{\psi}(\br,t) \right\rangle_0, \label{ap:j}
\end{align}
where $\braket{\cdots}_0$ denotes the expectation value in the quantum state, not random averaging.
Substituting the Fourier expansion of $\hat{\psi}(\br,t)$ and performing the integration of $\br$, we obtain
\begin{align}
\bj(t) = \int \frac{d^dk}{(2\pi)^d}  v_{\bk}  \left\langle \hat{\psi}^\dag_{\bk}(t)\hat{\psi}_{\bk}(t) \right\rangle_0,
\end{align}
where $\hat{\psi}_{\bk}(t)$ denotes the Fourier component of $\hat{\psi}(\br,t)$.
Here we define the lesser Green function as $G^<(\bk;t,t') \equiv i\left\langle \hat{\psi}^\dag_{\bk}(t')\hat{\psi}_{\bk}(t) \right\rangle_0$.
The Fourier expansion of $G^<(\bk;t,t')$ with respect to $t$ and $t'$ is
\begin{align}
&G^<(\bk;t,t') \notag\\
&=  \int_{-\infty}^{\infty} \frac{d\omega}{2\pi}  \int_{-\infty}^{\infty} \frac{d\omega'}{2\pi} \tilde{G}^<(\bk;\omega,\omega') 
e^{i\omega t} e^{-i\omega' t'}\\
&= \sum_{m,n}\int_{0}^{\Omega} \frac{d\omega}{2\pi} \int_{0}^{\Omega} \frac{d\omega'}{2\pi} \tilde{G}^<_{mn}(\bk;\omega,\omega') e^{i(\omega-m\Omega)t} e^{-i(\omega'-n\Omega)t'},
\end{align}
where we have used $\int_{-\infty}^{\infty} d\omega e^{i\omega t} = \sum_m \int_0^\Omega e^{i(\omega -m\Omega) t}$ and  $\tilde{G}^<_{mn}(\bk;\omega,\omega') \equiv \tilde{G}^<(\bk;\omega-m\Omega,\omega'-n\Omega)$ ($0 \leq \omega, \omega' \leq \Omega$).
For convenience, we extend the domain of $\tilde{G}^<_{mn}(\bk;\omega,\omega')$ by $\tilde{G}^<_{m+1,n+1}(\bk;\omega,\omega') = \tilde{G}^<_{mn}(\bk;\omega-\Omega,\omega'-\Omega)$.
We note that $m$ and $n$ represent the Floquet indices.

We assume $G^<(\bk;t,t') = G^<(\bk;t+T,t'+T) $,
meaning that the Green function is periodic with respect to the center of time $(t+t')/2$.
This assumption implies that we consider a nonequilibrium steady state represented by a mixed state of the Floquet eigenstates.
From this assumption, we have  $\tilde{G}^<_{mn}(\bk;\omega,\omega') = 2\pi \delta(\omega-\omega') G^<_{mn}(\bk,\omega)$,
and then the Green function at the same time $t=t'$ is given by
\begin{align}
G^<(\bk;t,t) 
&=\sum_{m,n} \int_{0}^{\Omega} \frac{d\omega}{2\pi} G^<_{mn}(\bk,\omega) e^{-i(m-n)\Omega t}.
\end{align}
Replacing the dummy variables of summation $(m,n)$ with $(l,n)$ ($l=m-n$)
and using the relation $G^<_{m+1,n+1}(\bk,\omega) = G^<_{mn}(\bk,\omega-\Omega)$,
we obtain
\begin{align}
G^<(\bk;t,t) 
&=\sum_{l} \int_{-\infty}^{\infty} \frac{d\omega}{2\pi} G^<_{l,0}(\bk,\omega) e^{-il\Omega t}.
\end{align}
As a result, the Fourier component of $\bj(t)$ at frequency $n\Omega$ is  obtained as
\begin{align}
\bj_n 
&= \int_0^T \frac{dt}{T} \bj(t) e^{in\Omega t} \notag \\
&=  -i  \int \frac{d^dk}{(2\pi)^d} \int_{-\infty}^\infty \frac{d\omega}{2\pi}v_{\bk} G^<_{n,0}(\bk,\omega). \label{ap:jn}
\end{align}
This is the formula by which we calculate the HHC from the non-equilibrium Green function.

%########################################################################
\section{Calculation of HHC}\label{sec:calcGF}
%#######################################################################
%########################################################################

%#### 
\begin{figure}[t]
\center
\includegraphics[width=7cm]{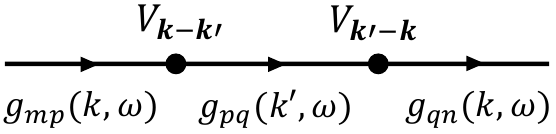}
\caption{
Feynmann diagram of the second-order corrections of the full Green functions $G^{(2)}$.
The solid lines and the dots indicate the bare Green function in the absence of the potential and the potential scattering respectively.
We integrate the internal momentum $\bk'$ and sum up the Floquet indices $p$ and $q$. 
}
\label{fig:fd}
\end{figure}
%#####

%#### 
\begin{figure*}
\center
\includegraphics[width=14cm]{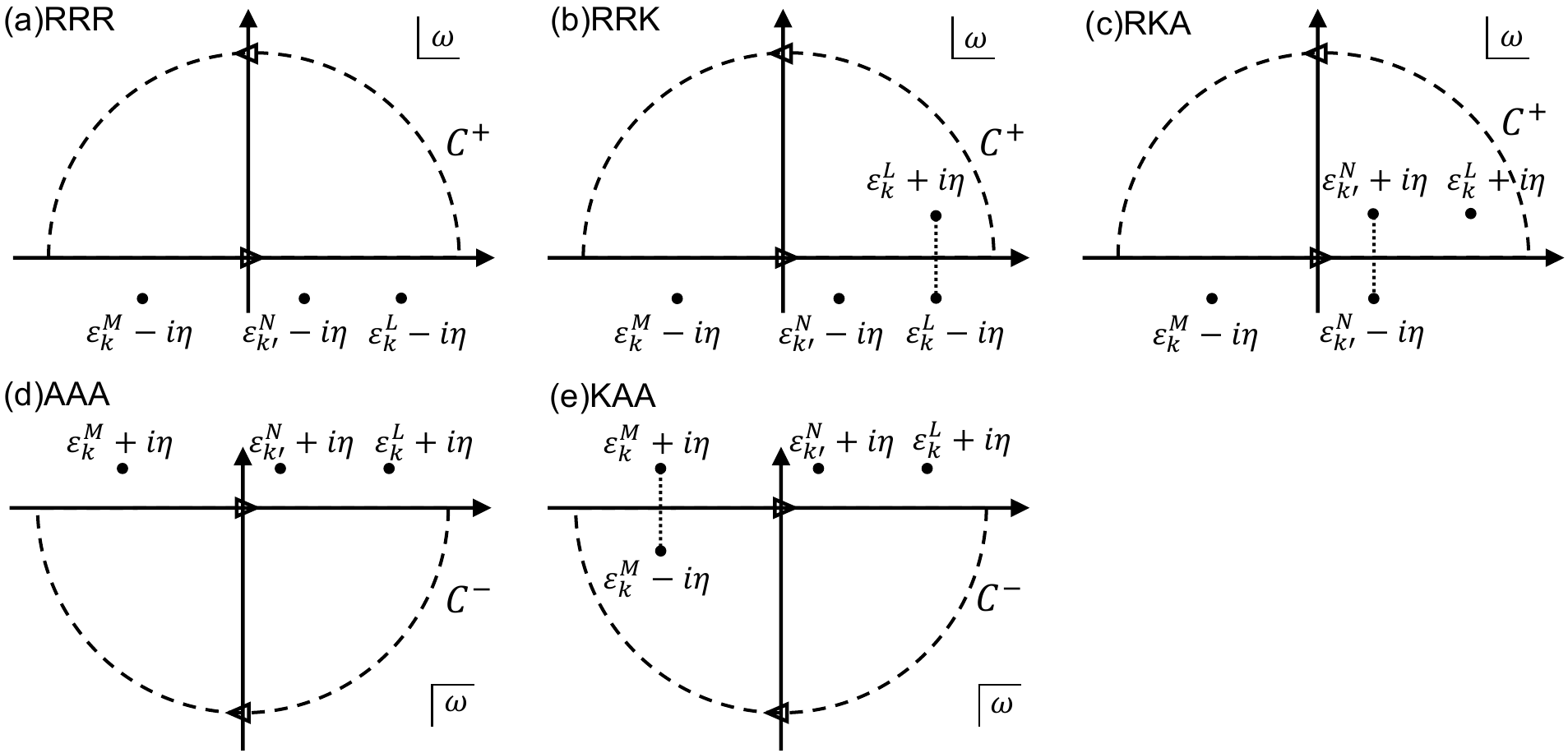}
\caption{Poles of (a) $RRR$, (b) $RRK$, (c) $RKA$, (d) $AAA$, and (e) $KAA$ terms in $\omega$-plane.
}
\label{fig:residue}
\end{figure*}
%#####

Here, we concretely calculate the $O(U^2)$ correction of the HHC
and derive Eqs.~\eqref{eq:jnbar}-\eqref{eq:fn}.
For convenience, we repeat the calculation procedure in the main text.
In the absence of the lattice potential, we obtain the exact bare Green function as follows,
\begin{align}
g_{mn}^{R/A}(\bk,\omega) 
&=\bra{\phi^m(\bk)} \frac{1}{\omega - \hf_0(\bk) \pm i\eta} \ket{\phi^n(\bk)} \notag \\
&= \sum_M \frac{ J_{M-m}(\alpha_{\bk}) J_{M-n}(\alpha_{\bk}) }{\omega - \epsilon_{\bk}^M \pm i \eta},\label{ap:gR} \\
 g_{mn}^{K}(\bk,\omega) 
&= -2 i \eta (1-2n_{\bk}) \sum_M  \frac{J_{M-m}(\alpha_{\bk}) J_{M-n}(\alpha_{\bk}) }{(\omega - \epsilon_{\bk}^M)^2 +  \eta^2}, \label{ap:gK}
\end{align}
where we have used  
$\sum_M \ket{\psi^M_0(\bk)}\bra{\psi^M_0(\bk)} = 1$, $\braket{\phi^m(\bk) | \psi^M_0(\bk)} = J_{M-m}(\alpha_{\bk})$, and $\hf_0(\bk) \ket{\psi^M_0(\bk)} = \epsilon_{\bk}^M \ket{\psi^M_0(\bk)}$.
In the presence of the potential, the exact full Green function is not available.
However, the lowest-order correction is obtained by the second-order Feynmann rules for the nonequilibrium Green function~\cite{Rammer1986},
\begin{align}
&G^{(2)}_{mn} (\bk,\omega) \notag\\
&= \sum_{p,q} \int \frac{d^d k'}{(2\pi)^d} g_{mp}(\bk,\omega) V_{\bk-\bk'} g_{pq}(\bk',\omega) V_{\bk'-\bk} g_{qn}(\bk,\omega), \label{ap:FeynmannRules}
\end{align}
where both $G_{mn}^{(2)}(\bk,\omega)$ and $g_{mn}(\bk,\omega)$ are the $2 \times 2$ matriices,
\begin{align}
G_{mn}^{(2)} (\bk,\omega) 
&= 
\begin{pmatrix}
G^R & G^K \\
0 & G^A
\end{pmatrix}_{mn}^{(2)}  (\bk,\omega),  \\
g_{mn} (\bk,\omega) 
&= 
\begin{pmatrix}
g^R & g^K \\
0 & g^A
\end{pmatrix}_{mn} (\bk,\omega).
\end{align}
Figure~\ref{fig:fd} shows the Feynmann diagram of $G^{(2)}$.
Once we obtain the Green function, we can calculate the high harmonics from Eq.~\eqref{ap:jn} with $G^<=(G^A-G^R+G^K)/2$.

Let us concretely calculate the second-order correction of the Green function from Eq.~\eqref{ap:gR}-\eqref{ap:FeynmannRules}.
From Eq.~\eqref{ap:FeynmannRules}, the retarded, the advanced and the Keldysh components are
\begin{align}
&G^{R(2)}(\bk,\omega)  = \int \frac{d^d k'}{(2\pi)^d} RRR , \\
&G^{A(2)}(\bk,\omega)  = \int \frac{d^d k'}{(2\pi)^d} AAA , \\
&G^{K(2)}(\bk,\omega) = \int \frac{d^d k'}{(2\pi)^d} ( RRK + KAA + RKA ),
\end{align}
where we have defined the following abbreviations:
\begin{align}
(XYZ)_{mn}
&\equiv  \sum_{p,q} g_{mp}^X(\bk,\omega) V_{\bk-\bk'} g_{pq}^Y(\bk',\omega) V_{\bk'-\bk} g_{qn}^Z(\bk,\omega) \notag\\
&\qquad \qquad  (X,Y, Z = R, K, A).
\end{align}
Therefore, the HHC at frequency $n\Omega$ is
\begin{align}
\bj_n 
&= -i \int \frac{d^d k}{(2\pi)^d} \int \frac{d\omega}{2\pi} v_{\bk} G^{<(2)}_{n0}(\bk,\omega)  \notag \\
&= -\frac{i}{2} \int \frac{d^d k}{(2\pi)^d} \int \frac{d^d k'}{(2\pi)^d} \int_{-\infty}^{\infty} \frac{d\omega}{2\pi} v_{\bk} \notag\\
&\qquad \times(AAA-RRR+RRK+KAA+RKA)_{n0}, \label{ap:1}
\end{align}
where we have used $G^< = (G^A-G^R +G^K)/2$.

We perform the $\omega$-integration in Eq.~\eqref{ap:1} by calculating the residues.
Since the integrand vanishes fast enough for $ |\omega| \rightarrow \infty$, 
we replace the integral path with the closed semicircle in the upper or  the lower halves of the $\omega$-plane.  
First, the $\omega$-integration of $AAA$ and $RRR$ terms vanish
because their integrands have poles only in the upper or the lower halves of the $\omega$-plane (see Figs. \ref{fig:residue}(a) and (d)).
Next, we perform the $\omega$-integrations of $RRK,KAA$ and $RKA$ terms.
From Eqs.~\eqref{ap:gR} and \eqref{ap:gK},  we calculate
\begin{widetext}
\begin{align}
\int_{C^+} \frac{d\omega}{2\pi} (RRK)_{n0}% \notag \\
&=-2i\eta (1-2n_{\bk})|V_{\bk-\bk'}|^2 \sum_{M,N,L,p,q}   \int_{C^+} \frac{d\omega}{2\pi}
\frac{ J_{M-n}(\alpha_{\bk}) J_{M-p}(\alpha_{\bk}) }{\omega - \epsilon_{\bk}^M + i \eta}
\frac{ J_{N-p}(\alpha_{\bk'}) J_{N-q}(\alpha_{\bk'}) }{\omega - \epsilon_{\bk'}^N + i \eta}
\frac{J_{L-q}(\alpha_{\bk}) J_{L}(\alpha_{\bk}) }{(\omega - \epsilon_{\bk}^L)^2 +  \eta^2} \notag \\
&=2\eta (1-2n_{\bk})|V_{\bk-\bk'}|^2 \sum_{M,N,L,p,q}  
\frac{ J_{M-n}(\alpha_{\bk}) J_{M-p}(\alpha_{\bk}) }{\epsilon_{\bk}^L - \epsilon_{\bk}^M + 2i \eta}
\frac{ J_{N-p}(\alpha_{\bk'}) J_{N-q}(\alpha_{\bk'}) }{\epsilon_{\bk}^L - \epsilon_{\bk'}^N + 2i \eta}
\frac{J_{L-q}(\alpha_{\bk}) J_{L}(\alpha_{\bk}) }{\epsilon_{\bk}^L - \epsilon_{\bk}^L +  2i\eta} \notag \\
&=-i (1-2n_{\bk})|V_{\bk-\bk'}|^2 \sum_{M,N,L,p,q} 
\frac{ J_{M-n}(\alpha_{\bk}) J_{M-p}(\alpha_{\bk}) J_{N-p}(\alpha_{\bk'}) J_{N-q}(\alpha_{\bk'}) J_{L-q}(\alpha_{\bk}) J_{L}(\alpha_{\bk}) }{ [(L -M)\Omega + 2i \eta] [(\epsilon_{\bk} - \epsilon_{\bk'}) + (L-N)\Omega + 2i \eta]},
\end{align}
\end{widetext}
where we have calculated  the residues owing to the second equality (see Fig.~\ref{fig:residue}(b)).
In the same way, we obtain
\begin{widetext}
\begin{align}
&\int \frac{d\omega}{2\pi} (KAA)_{n0} 
=i (1-2n_{\bk})|V_{\bk-\bk'}|^2 \sum_{M,N,L,p,q} 
\frac{ J_{M-n}(\alpha_{\bk}) J_{M-p}(\alpha_{\bk}) J_{N-p}(\alpha_{\bk'}) J_{N-q}(\alpha_{\bk'}) J_{L-q}(\alpha_{\bk}) J_{L}(\alpha_{\bk}) }{ [(L-M)\Omega + 2i \eta] [(\epsilon_{\bk} - \epsilon_{\bk'}) + (M-N)\Omega - 2i \eta]}, \\
&\int \frac{d\omega}{2\pi} (RKA)_{n0} 
= -i (1-2n_{\bk'})|V_{\bk-\bk'}|^2 \sum_{M,N,L,p,q} 
J_{M-n}(\alpha_{\bk}) J_{M-p}(\alpha_{\bk}) J_{N-p}(\alpha_{\bk'}) J_{N-q}(\alpha_{\bk'}) J_{L-q}(\alpha_{\bk}) J_{L}(\alpha_{\bk}) \notag \\
& \hspace{4cm} \times \left[
\frac{ (L-M)\Omega + 4i\eta }{ [(L-M)\Omega + 2i\eta] [(\epsilon_{\bk} - \epsilon_{\bk'}) + (L-N)\Omega + 2i \eta] [(\epsilon_{\bk} - \epsilon_{\bk'}) + (M-N)\Omega - 2i \eta] }
\right].
\end{align}
\end{widetext}
Figure~\ref{fig:residue} shows the poles of the each term in Eq.~\eqref{ap:1}.
Summing up all these terms, 
we obtain the following expression
\begin{widetext}
\begin{align}
&\int \frac{d\omega}{2\pi} (AAA - RRR + RRK + KAA + RKA)_{n0} \notag \\
&= -2i (n_{\bk}-n_{\bk'})|V_{\bk-\bk'}|^2 \sum_{M,N,L,p,q} 
J_{M-n}(\alpha_{\bk}) J_{M-p}(\alpha_{\bk}) J_{N-p}(\alpha_{\bk'}) J_{N-q}(\alpha_{\bk'}) J_{L-q}(\alpha_{\bk}) J_{L}(\alpha_{\bk}) \notag \\
& \hspace{2cm} \times \left[
\frac{ (L-M)\Omega + 4i\eta }{ [(L-M)\Omega + 2i\eta] [(\epsilon_{\bk} - \epsilon_{\bk'}) + (L-N)\Omega + 2i \eta] [(\epsilon_{\bk} - \epsilon_{\bk'}) + (M-N)\Omega - 2i \eta] }
\right] \notag \\
&= 2i \sum_l \frac{(n_{\bk} - n_{\bk'}) |V_{\bk-\bk'}|^2 (n\Omega - 4i\eta) J_{-l}(\alpha_{\bk-\bk'}) J_{n+l}(\alpha_{\bk'-\bk})}{\left(n\Omega - 2i\eta \right)  \left[(\epsilon_{\bk} - \epsilon_{\bk'}) - l\Omega - 2i\eta \right]  \left[(\epsilon_{\bk'} - \epsilon_{\bk}) + (n+l)\Omega - 2i\eta \right] } \notag \\
&= 2i \sum_{a,b} f_n(\bk,\bk') e^{-i(\bk-\bk')\cdot \br_a} e^{i(\bk-\bk')\cdot \br_b}, \label{ap:2} \\
&f_n(\bk,\bk') 
=  U^2 |u_{\bk-\bk'}|^2 (n_{\bk} - n_{\bk'}) \left( \frac{n\Omega - 4i\eta}{n\Omega - 2i\eta} \right)
\sum_l \frac{ J_{-l}(\alpha_{\bk-\bk'}) J_{n+l}(\alpha_{\bk'-\bk})}{ \left[(\epsilon_{\bk} - \epsilon_{\bk'}) - l\Omega - 2i\eta \right] \left[(\epsilon_{\bk'} - \epsilon_{\bk}) + (n+l)\Omega - 2i\eta \right] }. \label{ap:fn}
\end{align}
\end{widetext}
where we have used $\sum_a J_{a}(x)J_{b-a}(y) = J_b(x+y)$ and $V_{\bk} = U \sum_a u_{\bk} e^{-i \bk \cdot \br_a}$.
Taking the average over the random variables $\delta \br_a$'s [see Eq.~\eqref{eq:phase_avg} in the main text] and substituting Eq.~\eqref{ap:2} and \eqref{ap:fn} to Eq.~\eqref{ap:1}, we obtain Eqs.~\eqref{eq:jnbar}--\eqref{eq:fn} in the main text.

%################################################################
%################################################################
\section{Relaxation-time insensitivity of the results}
%#### 
\begin{figure}[t]
\center
\includegraphics[width=8cm]{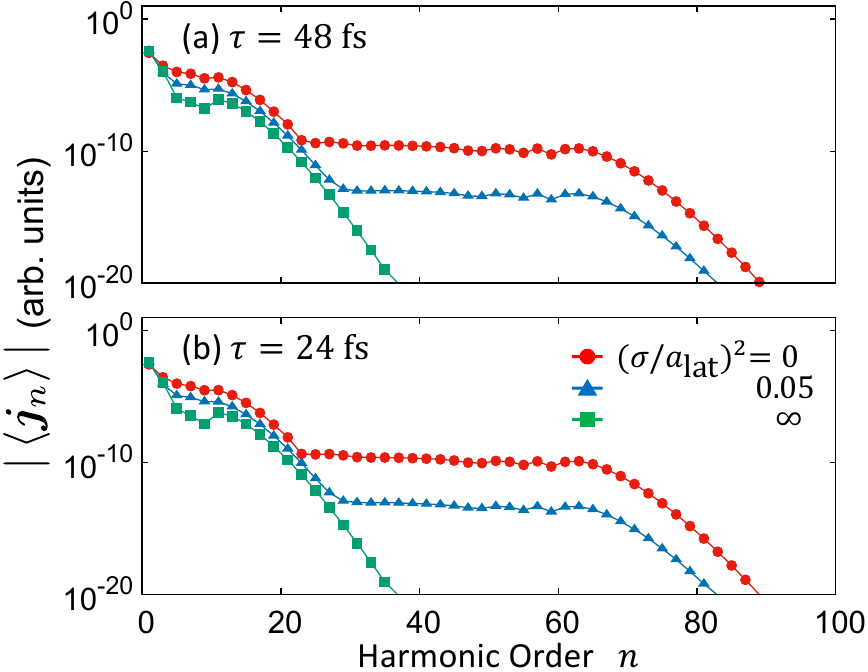}
\caption{
Harmonic spectra for the relaxation times $\tau = 48$\,$\mathrm{fs}$ (a) and $24$\,$\mathrm{fs}$ (b).
In each panel, we plot $| \langle\bj_n\rangle |$ for $(\sigma/a_\text{lat})^2=0$ (circle), 0.05 (triangle), and $\infty$ (square).
The laser field is $E_0 =A_0\Omega=10.6$\,$\mathrm{MV/cm}$.
}
\label{fig:tau}
\end{figure}
%#####

Figure~\ref{fig:tau} shows the harmonic spectra for two relaxation times $\tau = 48$\,$\mathrm{fs}$ and $24$\,$\mathrm{fs}$.
The panel (a) is the same as Fig.~3(a) in the main text,
and the panel (b) is calculated with $\tau$ changed and all the other parameters fixed.
Evidently, the harmonic spectra are almost the same.
Thus we confirm that the physical consequences are not sensitive to 
the relaxation time.

%\end{widetext}

%\bibliography{../../../../../bibtex/random_hhg_project}%#### for chinzei
%\bibliography{../../../../../../Bibtex/random_hhg_project}%### for ikeda 


\end{document}